\documentclass[published]{nst}

\usepackage{subfigure,dcolumn}
\usepackage[T2A,T1]{fontenc}
\usepackage[english]{babel}

\usepackage{listings}
\begin{document}

\title{Design of High-energy Proton-beam Experiment Station at CSNS}

\author{Yu-Hang Guo}
\affiliation{Institute of High Energy Physics,  Chinese Academy of Sciences (CAS), Beijing 100049, China}
\affiliation{Spallation Neutron Source Sciences Center, Dongguan 523803, China}

\author{Han-Tao Jing}
\email[Corresponding author, ]{jinght@ihep.ac.cn}
\affiliation{Institute of High Energy Physics,  Chinese Academy of Sciences (CAS), Beijing 100049, China}
\affiliation{Spallation Neutron Source Sciences Center, Dongguan 523803, China}

\author{Ming-Yi Dong}
\email[Corresponding author, ]{dongmy@ihep.ac.cn}
\affiliation{Institute of High Energy Physics,  Chinese Academy of Sciences (CAS), Beijing 100049, China}

\author{Zhi-Ping Li}
\affiliation{Institute of High Energy Physics,  Chinese Academy of Sciences (CAS), Beijing 100049, China}
\affiliation{Spallation Neutron Source Sciences Center, Dongguan 523803, China}
\author{Yong-Ji Yu}
\affiliation{Institute of High Energy Physics,  Chinese Academy of Sciences (CAS), Beijing 100049, China}
\affiliation{Spallation Neutron Source Sciences Center, Dongguan 523803, China}
\author{Yan-Liang Han}
\affiliation{Institute of High Energy Physics,  Chinese Academy of Sciences (CAS), Beijing 100049, China}
\affiliation{Spallation Neutron Source Sciences Center, Dongguan 523803, China}
\author{Zhi-Xin Tan}
\affiliation{Institute of High Energy Physics,  Chinese Academy of Sciences (CAS), Beijing 100049, China}
\affiliation{Spallation Neutron Source Sciences Center, Dongguan 523803, China}
\author{Zhi-Jun Liang}

\author{Sen Qian}
\affiliation{Institute of High Energy Physics,  Chinese Academy of Sciences (CAS), Beijing 100049, China}
\author{Hong-Yu Zhang}
\affiliation{Institute of High Energy Physics,  Chinese Academy of Sciences (CAS), Beijing 100049, China}
\author{Han Yi}
\affiliation{Institute of High Energy Physics,  Chinese Academy of Sciences (CAS), Beijing 100049, China}
\affiliation{Spallation Neutron Source Sciences Center, Dongguan 523803, China}
\author{You Lv}
\affiliation{Institute of High Energy Physics,  Chinese Academy of Sciences (CAS), Beijing 100049, China}
\affiliation{Spallation Neutron Source Sciences Center, Dongguan 523803, China}
\author{Qiang Li}
\affiliation{Institute of High Energy Physics,  Chinese Academy of Sciences (CAS), Beijing 100049, China}
\affiliation{Spallation Neutron Source Sciences Center, Dongguan 523803, China}
\author{Xin Shi}
\affiliation{Institute of High Energy Physics,  Chinese Academy of Sciences (CAS), Beijing 100049, China}
\affiliation{Spallation Neutron Source Sciences Center, Dongguan 523803, China}
\author{Xiao-Fei Gu}
\affiliation{Zhengzhou University, Zhengzhou 450001, China}
\author{Yi Liu}
\affiliation{Zhengzhou University, Zhengzhou 450001, China}
\author{Xiu-Xia Cao}
\affiliation{Institute of High Energy Physics,  Chinese Academy of Sciences (CAS), Beijing 100049, China}
\affiliation{Spallation Neutron Source Sciences Center, Dongguan 523803, China}
\author{Si-Xuan Zhuang}
\affiliation{Institute of High Energy Physics,  Chinese Academy of Sciences (CAS), Beijing 100049, China}
\affiliation{Spallation Neutron Source Sciences Center, Dongguan 523803, China}
\author{Lan-Kun Li}
\affiliation{Institute of High Energy Physics,  Chinese Academy of Sciences (CAS), Beijing 100049, China}
\author{Meng-Zhao Li}
\affiliation{Institute of High Energy Physics,  Chinese Academy of Sciences (CAS), Beijing 100049, China}
\affiliation{Spallation Neutron Source Sciences Center, Dongguan 523803, China}
\author{Yun-Yun Fan}
\affiliation{Institute of High Energy Physics,  Chinese Academy of Sciences (CAS), Beijing 100049, China}
\author{Hao He}
\affiliation{Institute of High Energy Physics,  Chinese Academy of Sciences (CAS), Beijing 100049, China}
\author{Li-Shuang Ma}
\affiliation{Institute of High Energy Physics,  Chinese Academy of Sciences (CAS), Beijing 100049, China}
\author{Rui-Rui Fan}
\affiliation{Institute of High Energy Physics,  Chinese Academy of Sciences (CAS), Beijing 100049, China}
\affiliation{Spallation Neutron Source Sciences Center, Dongguan 523803, China}
\author{Zhi-Jia Sun}
\affiliation{Institute of High Energy Physics,  Chinese Academy of Sciences (CAS), Beijing 100049, China}
\affiliation{Spallation Neutron Source Sciences Center, Dongguan 523803, China}
\author{Yuan-Bo Chen}
\affiliation{Institute of High Energy Physics,  Chinese Academy of Sciences (CAS), Beijing 100049, China}
\affiliation{Spallation Neutron Source Sciences Center, Dongguan 523803, China}

\begin{abstract}
China's first proton test beam facility, named the High-energy Proton-beam Experiment Station (HPES), is currently under construction in campus of CSNS, as part of the CSNS-II project. 
Utilizing protons slowly extracted from the Rapid Cycling Synchrotron of CSNS, HPES will deliver 1.6 GeV proton beam with an adjustable flux ranging from 1E3 to 1E8 protons per second. 
The station is composed of two dedicated test terminals designed to support comprehensive beam tests, serving as an advanced platform for particle detector development, irradiation hardness studies of aerospace chips, and GeV-proton-induced nuclear data measurements.
To characterize the beam, HPES incorporates dedicated flux and profile monitors. 
For user experiments, the facility is equipped with a high-precision proton telescope offering a positioning resolution of 10 $\mu$m, and a Time-of-Flight (TOF) spectrometer achieving an energy resolution of 1\%. 
Furthermore, a compatible trigger logic unit have been designed to provide precise event tagging, which is essential for data alignment. 
This paper presents an overview of the detector systems within HPES, discusses their design considerations, and outlines the future prospects of the facility.
\end{abstract}


\maketitle

\section{Introduction}\label{sec:intro}

High-energy particle colliders \cite{ATLAS,CMS,CEPCTDR,FCC,STCF} are indispensable tools for exploring the fundamental laws of physics. 
Within these colliders, a vast array of particle detectors, including silicon pixel and strip detectors, time-projection chambers (TPCs), drift chambers, and scintillators coupled with silicon photomultipliers (SiPMs), act as the primary instruments for detecting secondary particles generated by collisions, which carry crucial physical information \cite{CALOR,TRACK}. 
The scale of detector integration in modern colliders is immense. 
For instance, the CMS detector \cite{CMS} comprises over 1,400 pixel detectors and 15,000 strip detectors forming its tracker, along with approximately 68,000 PbWO$_4$ crystals and 70,000 plastic scintillators constituting its calorimeter system. 
Constructing such complex detector systems requires rigorously designed detection technologies and precisely calibrated integration. 
This is the motivation that drives the necessity for dedicated test beam facilities.

\begin{figure}
	\centering
	\includegraphics[width=.9\hsize]{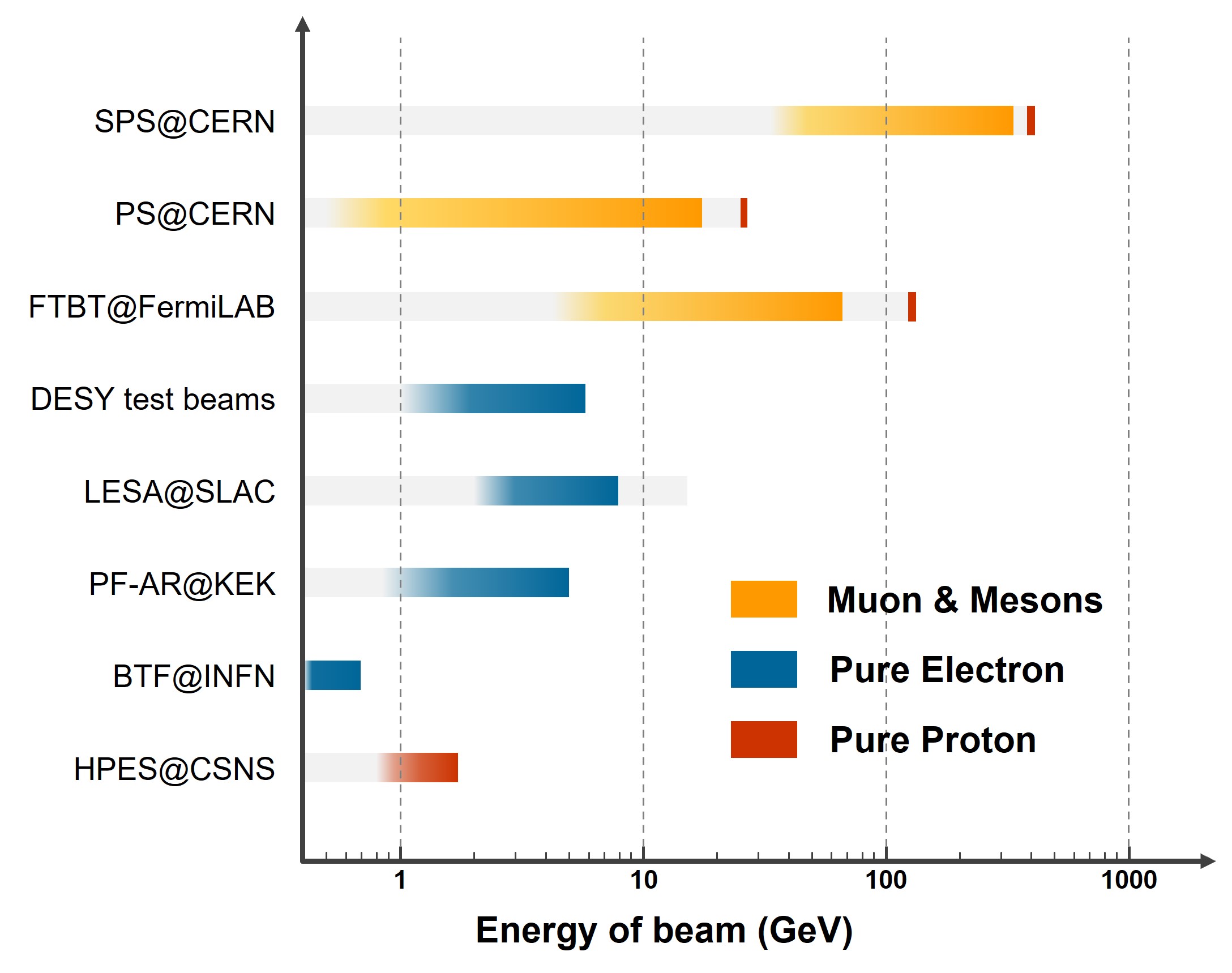}
	\caption{Energy range of representative test beams. HPES poses a valuable supplement in GeV range.}
	\label{fig:tb}
\end{figure}

Test beams typically provide high-energy charged particle beams to evaluate the performance of particle detectors. 
When the beam energy corresponds to the Minimum Ionizing Particle (MIP) regime, the penetration depth is maximized, making it ideal for assessing tracking detectors and measuring spatial resolution. 
Furthermore, high-energy beams can penetrate the inner layers of multi-layered calorimeters, enabling comprehensive performance validation. 
Numerous test beam facilities have been established worldwide. 
Fig. \ref{fig:tb} has provided an basic comparison among the test beams running currently.
Notable examples include the facility operated by Deutsches Elektronen-Synchrotron (DESY) \cite{DESY}, which provides electron beams in the 1--6~GeV energy range. 
At CERN, test beams are available in both the North and East Areas \cite{CERN-north,CERN-east-2019,CERN-east-2021}.
The North Area offers pure proton beam or mixed secondary hadron beams with momenta up to 400~GeV/$c$, while the East Area provides pure or mixed electron, muon, and hadron beams with momenta up to 16~GeV/$c$. 
In the United States, the FTBT \cite{FTBF} at Fermilab delivers protons at 120~GeV and secondary particles ranging from $\sim$200~MeV to 60~GeV, while the LESA \cite{LESA} at SLAC provides electron beams up to 8 GeV as a successor of the ESTB facility \cite{ESTB}. 
Additionally, electron test beams are available at INFN and KEK for user experiments\cite{INFN,KEK}.

Recently, a new proton test beam facility, named the High-energy Proton-beam Experiment Station (HPES), is under construction at the China Spallation Neutron Source (CSNS) \cite{CSNS-NAT,CSNS} campus in Dongguan, China. 
As part of the CSNS upgrade project (CSNS-II), HPES is scheduled to deliver its first beam by the end of 2029. 
It is designed to offer a flexible proton flux ranging from $10^3$ to $10^8$ protons per second. 
The high-flux mode is intended for studying radiation effects on chips and electronic boards, whereas the low-flux mode, approaching a "single proton per pulse" regime, is tailored for particle detector characterization. 
Simulation studies regarding the generation of this 1.6~GeV proton beam have been conducted by Zhou Kai, et al. \cite{RABT}. Key beam properties will be briefly discussed in Chapter~\ref{sec:beam}.

To facilitate user experiments, seven detection Devices In Terminals (DITs) have been developed, including a proton beam telescope based on Silicon Pixel detectors \cite{SiPix1,SiPix2} and a proton energy spectrometer utilizing Low Gain Avalanche Diodes (LGADs) \cite{LGAD1,LGAD2,LGAD-irr-0,LGAD-1,LGAD-2}. 
A general overview of these devices is provided in Table~\ref{tbl:7detector}, with detailed descriptions presented in Chapter~\ref{sec:device}. 
For tests involving particle trackers and calorimeters, precise data alignment among devices is critical for analysis. 
Consequently, a Trigger Logic Unit (TLU) has been developed based on the widely adopted AIDA-2020 architecture \cite{AIDA2020}. 
While maintaining compatibility with the AIDA standard, specific improvements have been implemented to accommodate the unique time structure of the CSNS beam. 
The architecture of the HPES TLU and the associated data control flow are detailed in Chapter~\ref{sec:tlu}. 
Finally, the artical concludes with a summary and a brief outlook on the anticipated applications and beam test campaigns at HPES.

\begin{table}[]
\caption{Parameters of the seven detection devices in HPES.}\label{tbl:7detector}
\begin{tabular*}{8cm} {@{\extracolsep{\fill} } cc}
\toprule
Name & Description \\
\midrule
HEPTel &  Proton telescope \\
LEMS &  Proton energy spectrometer \\
FLASH &  Proton trigger Device  \\
PALET &  Beam profile detector \\
PROUD &  Beam Tuning detector \\
SEEM &  Beam flux online monitor \\
BMOS &  Beam flux online monitor \\
\bottomrule
\end{tabular*}
\end{table}

\section{Facility Layout and Beam Parameters}\label{sec:beam}

\subsection{Layout of HPES}

\begin{figure}
    \centering
    \subfigure[Overview of high energy proton test area]{    
		\label{fig:map_a}     
	    \includegraphics[width=.8\hsize]{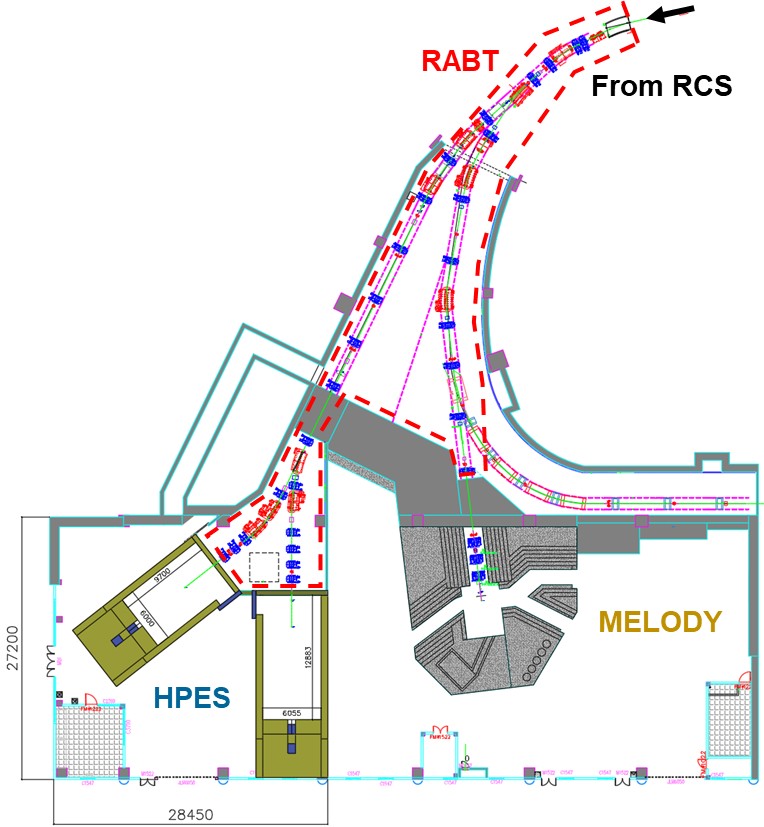}}
    \qquad
    \subfigure[Zoomed-in view of HPES hall]{    
		\label{fig:map_b}     
		\includegraphics[width=.9\hsize]{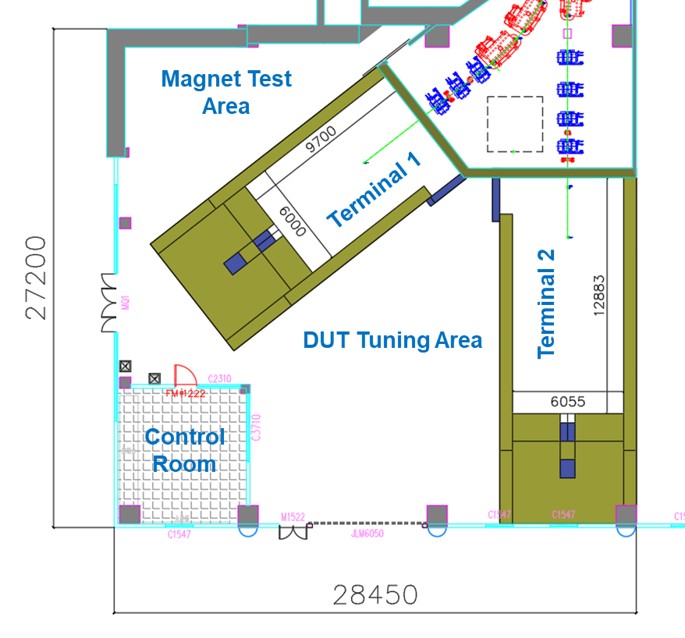}}
	\caption{Layout of HPES. Two terminals have been designed.}
	\label{fig:map}
\end{figure}

Figure~\ref{fig:map} illustrates the layout of HPES. 
Located within the high-energy proton test area of CSNS campus, the HPES is adjacent to the RABT beamline \cite{RABT} and MELODY \cite{MELODY} which is to be China's first surface muon source. 
The HPES hall houses two dedicated experimental terminals designed for user experiments. 
Radiation safety is ensured by concrete shielding walls, interlocked shielding doors, and beam dumps located at the end of each terminal. These measures attenuate the radiation dose rate outside the experimental areas to below 2.5~$\mu$Sv/h, allowing personnel to remain safely in the adjacent control zones during beam delivery. 
The central region between the two terminals serves as a Device Under Test (DUT) preparation area, where users can configure and tune their setups prior to beam exposure. 
Additionally, space in the upper corner of the HPES hall is reserved for testing high-field magnets, which will support experiments requiring specific magnetic field environments. 
The two terminals operate in an alternating mode. 
This strategy will provide opportunity for the \textit{in-situ} integration and DUT testing in one terminal while the other is active.

\subsection{Beam Parameters of HPES}

The proton beam delivered to HPES is generated via the "scattering slow extraction" method, utilizing a rotating carbon scattering foil inserted into the intensive main proton beam runing in the Rapid Cycling Synchrotron (RCS) of CSNS. 
The RCS operates with a repetition rate of 25~Hz. 
In each cycle, an 80~MeV proton beam from the CSNS LINAC is accumulated and accelerated to 1.6~GeV over approximately 19~ms. 
The beam is then maintained at energy of 1.6~GeV for 1~ms before it is extracted to the CSNS target station for spallation neutron production. 
Synchronized with the RCS cycle, the carbon foil rotates at 25~Hz, intercepting the beam halo during the 1~ms energy-maintaining period. 
The total number of protons circulating in the RCS per cycle is approximately $2.89\times 10^{13}$. 
A small fraction of these protons undergoes elastic scattering at angles between $20-25$~mrad and be extracted by a Lamberson Magnet, subsequently exits the ring and transport to the HPES terminals via the RABT beamline.

\begin{figure}
	\centering
	\includegraphics[width=.9\hsize]{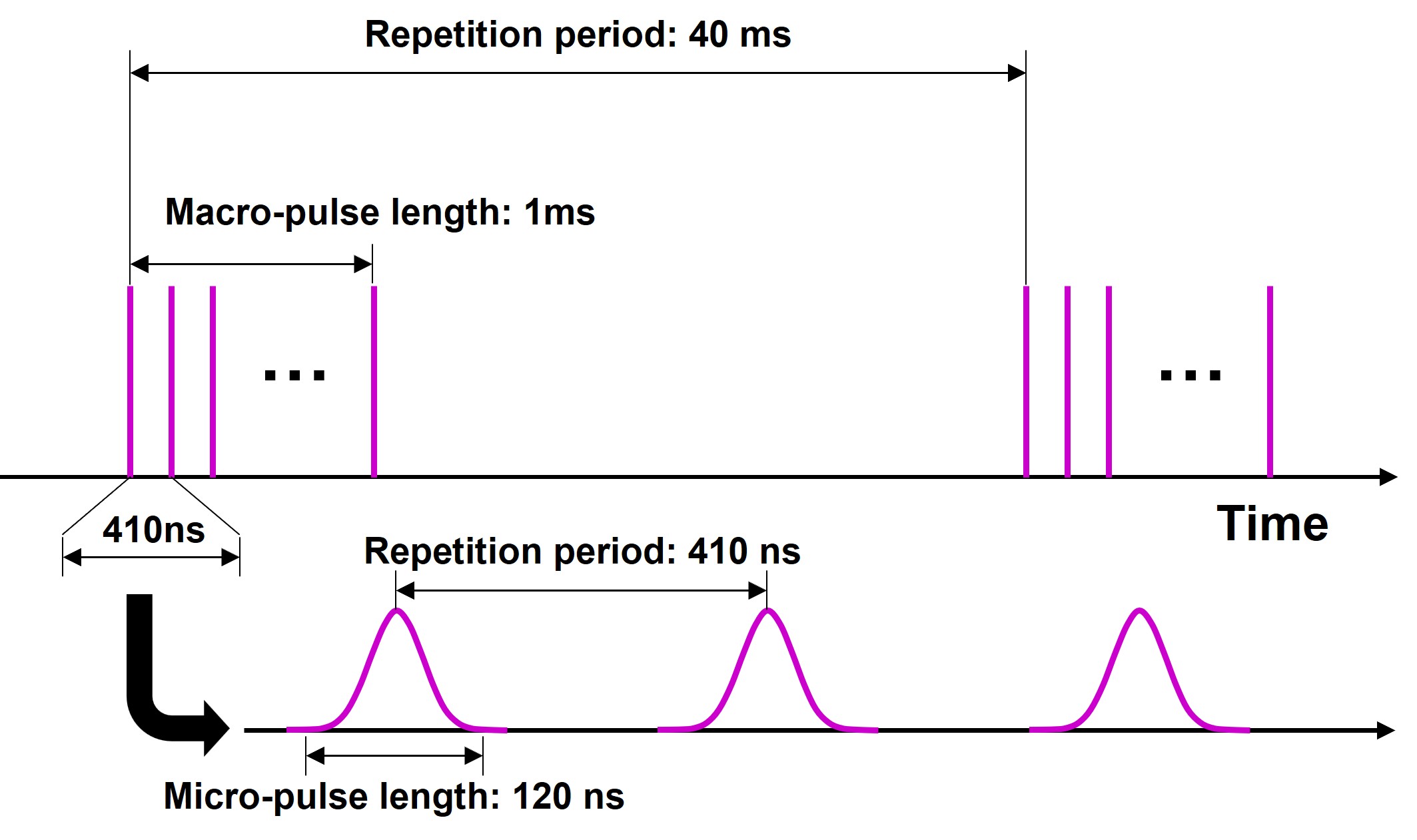}
	\caption{Temporal structure of the proton beam in HPES. It is composed of a dual-layer hierarchy of macro-pulses and micro-pulses. Not to scale.}
	\label{fig:ts}
\end{figure}

As illustrated in Fig.~\ref{fig:ts}, the temporal structure of the proton beam features a dual-layer pulse hierarchy. 
The macro-pulse repeats at 25~Hz with a duration of 1~ms, dictated by the RCS cycle. 
However, one of the the 25 macro-pulses in every second is blank since the main beam in this cycle will be delivered to the muon source. 
Thus, the effective frequency of the macro-pulse in HPES is 24 Hz.
Within each macro-pulse, the beam interacts with the scattering foil every 410~ns (the revolution period of the RCS), generating a micro-pulse structure. 
Consequently, the HPES beam can be characterized as a pulsed beam with a 25~Hz macro-repetition rate, containing a quasi-continuous train of micro-pulses during each macro-pulse.

\begin{figure}[h]
	\centering
	\includegraphics[width=.9\hsize]{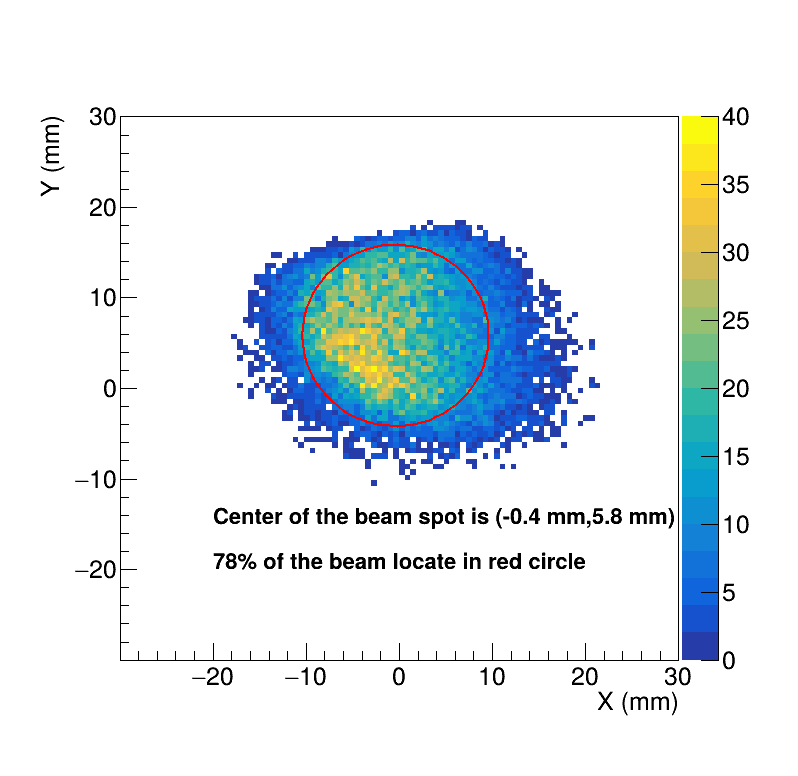}
	\caption{Simulated profile of beam spot at test point in terminals of HPES. Center of the red circle in the plot locates in the center of beam spot. Radius is 10 cm. 78\% of the beam will distribute in the red circle.}
	\label{fig:beamspot}
\end{figure}

The RABT beamline comprises a series of dipole and quadrupole magnets, two collimators, a degrader, and associated auxiliary systems. 
Dipole magnets steer the beam toward the designated terminals, while quadrupole magnets focus the beam to minimize losses due to dispersion. 
Through precise magnet tuning, the beam spot profile at the experimental test point is constrained to $2\times 2$~cm$^2$ as illustrated in Fig. \ref{fig:beamspot}, which is a dimension optimized for tracking detector characterization. 
The dimension of beam spot can also be tuned to $10\times 10$~cm$^2$ at most, according to the requirement of the users.
The proton flux at HPES is continuously adjustable over a range of $10^3$--$10^8$~protons per second. 
Primary flux control is achieved by varying the insertion depth of the scattering foil into the main RCS beam. 
Simulation studies using \texttt{G4PyOrbit} \cite{G4PyORBIT} confirm that deeper insertion yields higher extracted flux. 
Fine-tuning of the flux is further accomplished using the collimators along the RABT. 
By combining the shallowest foil insertion with the smallest collimator aperture, the beam intensity can be reduced to the single-particle-per-pulse regime, a condition highly preferred for precision particle detector tests.

\begin{figure}[h]
	\centering
	\includegraphics[width=.9\hsize]{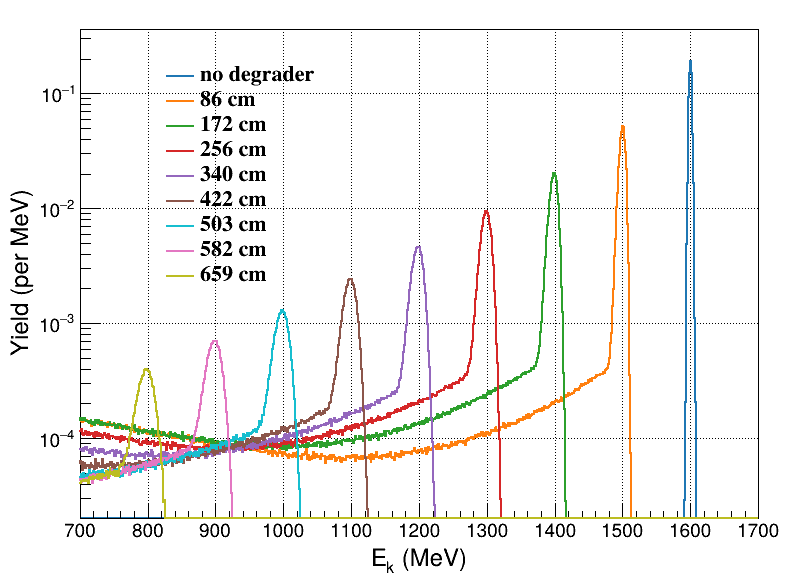}
	\caption{Simulated proton flux with different traversing length in degrader.}
	\label{fig:energy}
\end{figure}

The energy of proton beam is primarily 1.6 GeV, while the spread of energy is less than 2.5\%.
The proton beam energy is also tunable via the installation of a degrader. 
This device consists of a tapered iron block positioned transversely across the beam path. 
By changing the transverse position of the degrader, the length of the beam traversing the degrader can be varied.
As illustrated in Fig. \ref{fig:energy}, a greater traversing length results in increased energy loss, thereby reducing the exit proton energy. 
Through precise adjustment of the traversing length, the proton beam energy can be continuously degraded from the nominal 1.6~GeV down to a minimum of 800~MeV. 
This lower limit is determined by optimizing the balance between the desired energy reduction and the preservation of sufficient beam flux, as excessive degradation leads to significant intensity loss and emittance growth.

\begin{table}
\caption{Parameters of beam parameters in HPES. Details see text.}\label{tbl:beampars}
\begin{tabular*}{8cm} {@{\extracolsep{\fill} } cc}
\toprule
Parameters & Designed Values \\
\midrule
Primary beam energy &  1.6 GeV  \\
Energy spread &  $<$2\% @ 1.6 GeV  \\
Tunable energy range & $0.8-1.6$ GeV \\
Proton flux &  $10^3-10^8$ p/s  \\
Macro-pulse frequency &  25 Hz  \\
Effective macro-pulse frequency & 24 Hz \\
Macro-pulse length &  1 ms  \\
Interval of micro-pulse &  410 ns \\
\bottomrule
\end{tabular*}
\end{table}

\section{Detection Devices}\label{sec:device}

\subsection{Proton Telescope}

In the high-energy particle colliders, the spatial resolution of tracking detectors is a critical parameter that decides the precision of momentum reconstruction and vertex identification. 
To ensure the performance of these detectors, a deep understanding to the performance in the development period and precise calibrations to the integrated ladders in the installation period using a reference trajectory is required before their deployment. 
At the HPES, a EUDET-type beam telescope \cite{EUDET}, named of High Energy Proton-beam Telescope (HEPTel), has been equipped to HPES.
By reconstructing the passage of incident particles with superior accuracy, the telescope provides a benchmark against which the position measurements of DUT are compared, allowing for the precise determination of the DUT's intrinsic spatial resolution and alignment.

\begin{figure}[h]
	\centering
	\includegraphics[width=.9\hsize]{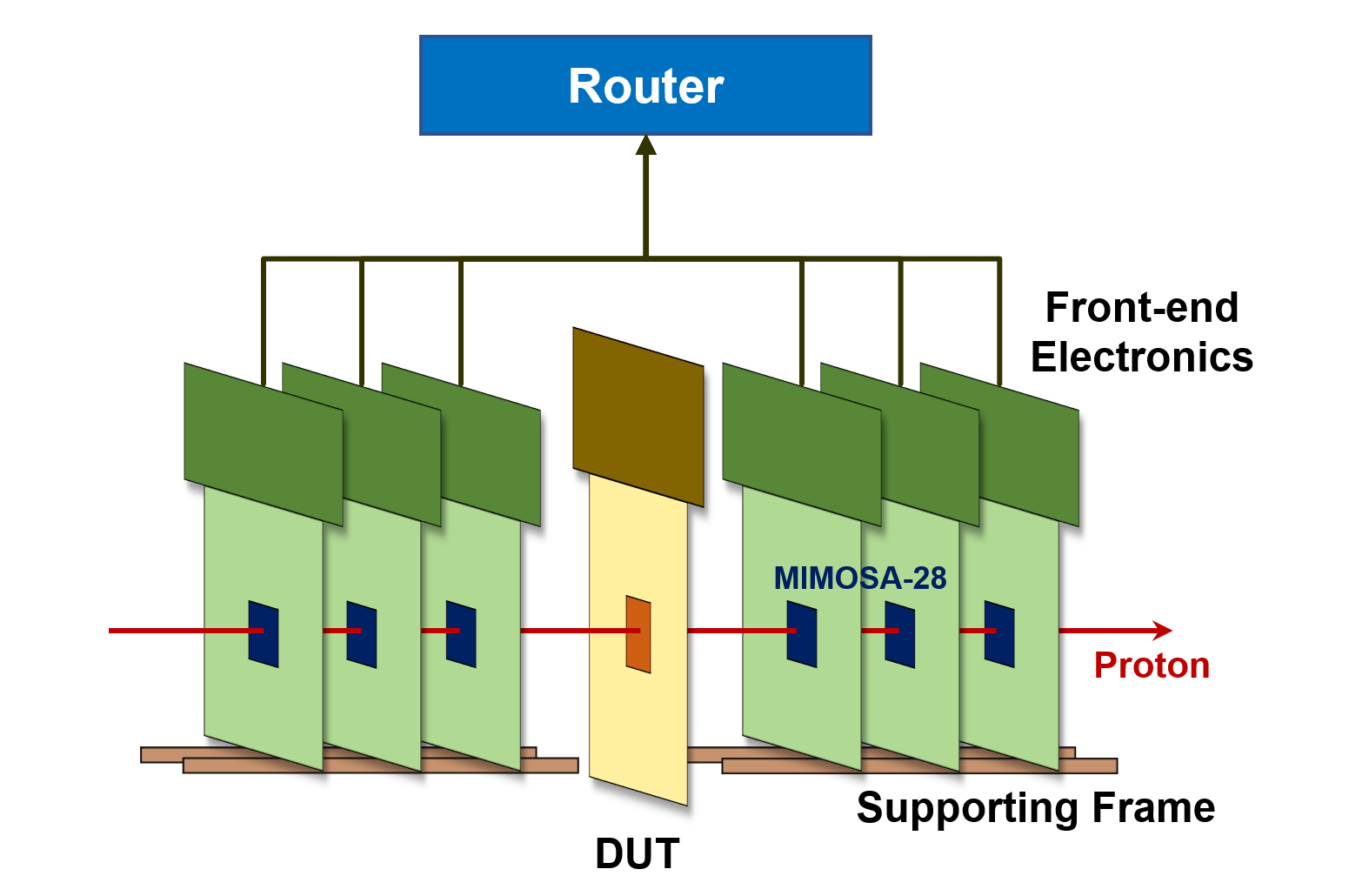}
	\caption{Diagram of the HEPTel. Six MIMOSA-28 pixel detectors have been applied to reconstruct the proton track and comparison to the positioning result of DUT would give the positioning ability of the DUT.}
	\label{fig:HEPTel-method}
\end{figure}

\begin{figure}[b]
	\centering
	\includegraphics[width=1.\hsize]{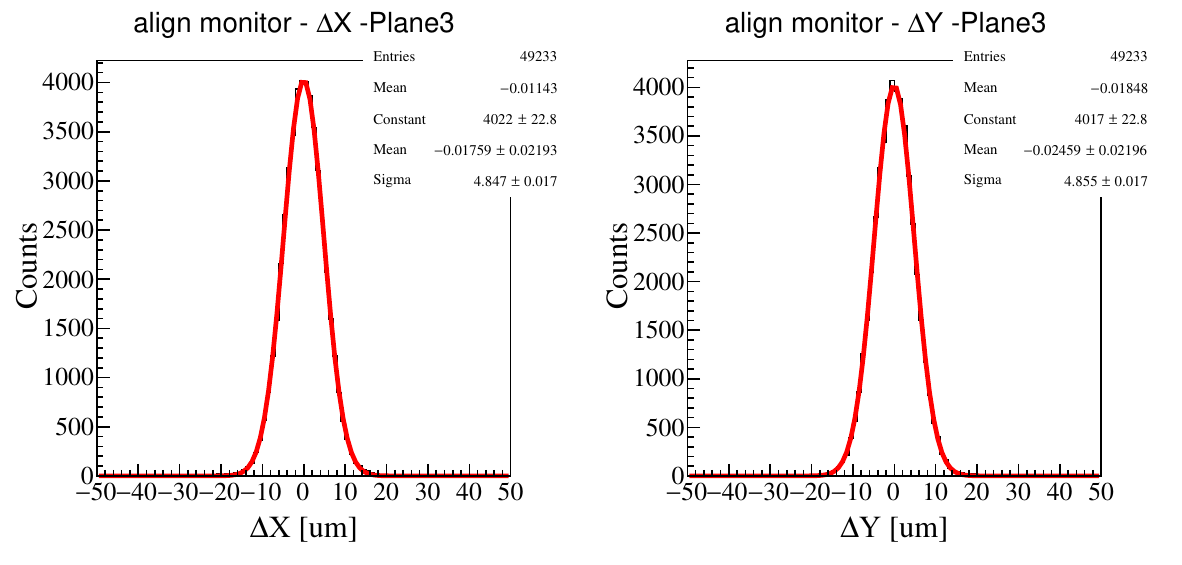}
	\caption{Reconstructed DUT resolution of a MIMOSA-28 detector in X (left panel) and Y (right panel) axis by HEPTel. The data is acquired based on the simulation study with 1.6 GeV injected proton beam. The DUT gap and Telescope gap are 2 cm. Details see Ref. \cite{HEPTel-res}}
	\label{fig:HEPTel-reso}
\end{figure}

As illustrated in Fig. \ref{fig:HEPTel-method}, the core components of HEPTel are silicon pixel detectors, chosen for their exceptional position resolution and low material budget, which minimizes multiple Coulomb scattering and preserves the accuracy of the reconstructed track. 
The system employs the MIMOSA-28 sensor \cite{MIMOSA28}, a state-of-the-art Monolithic Active Pixel Sensor (MAPS) \cite{MAPS}. 
The MIMOSA-28 features a pixel pitch of 20.7~mum and an ultra-thin active thickness of only 50~$\mu $m. 
Based on operational experience at international facilities such as DESY, a beam telescope constructed with MIMOSA sensors can achieve a track reconstruction precision as high as 1.30~mum \cite{EUDET-res}.
However, this performance is hard to be reached since the beam energy of HPES is not large enough, which will cause relatively significant multi-scatterings and worsen the resolution track reconstruction \cite{HEPTel}.
As illustrated in Fig. \ref{fig:HEPTel-reso}, simulation studies has been conducted with the \texttt{Allpix}\textsuperscript{\texttt{2}} \cite{ALLPIX}. 
The result shows, with the 1.6 GeV proton beam, the resolution of calibration to DUT can only reach level of ~4.8 $\mu$m in both X and Y axis, whereas the telescope resolution for tracking reaches 1.83 $\mu$m \cite{HEPTel-res}. 
However, this performance is enough for most of the tracking detectors. 

\subsection{Proton Energy Spectrometer}

Another critical test requirement from colliders is the test to calorimeters. 
To support this, the HPES provides a proton beam with a continuously tunable energy range of 0.8 to 1.6 GeV, enabling the calibration and performance understanding of hadron calorimeter in this energy range. 
However, when the beam energy is reduced using the degrader, the proton energy spectrum undergoes significant broadening. 
Consequently, to ensure precise calibration, it is necessary to determine the energy of each individual proton on an event-by-event basis. 
This necessitates the design of a energy measurement device to correlate the incident energy with the calorimeter response. which makes minimum influence to the energy of protons. 
To fulfill this requirement, the HPES employs a proton energy spectrometer, named as LGAD based Energy Measurement System (LEMS), based on Low Gain Avalanche Detectors (LGAD) utilizing the Time-of-Flight (TOF) technique.

\begin{figure}
	\centering
	\includegraphics[width=.9\hsize]{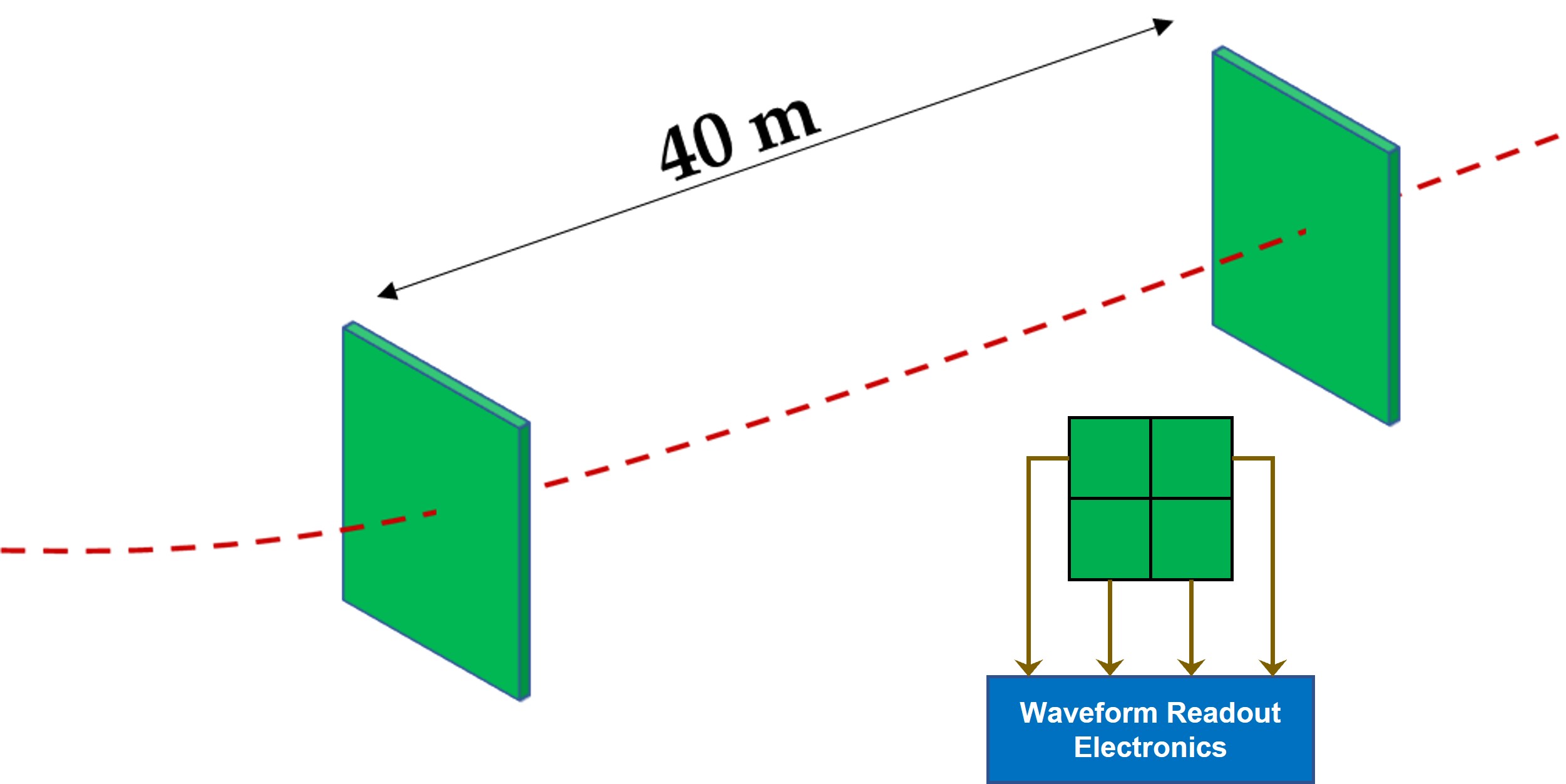}
	\caption{The schematic of LEMS. Two LGAD arrays with $15\times 15$ segmentation cells are placed with a distance of 40 m to each other. The segmentation cells are bonded into 4 readout sectors.}
	\label{fig:LEMS}
\end{figure}

LGADs are semiconductor detectors characterized by a low material budget and exceptional timing capabilities, with intrinsic time resolutions reaching up to 30 ps and a thickness of only 300 $\mu $m. 
These features make it ideally suited for measuring the kinetic energy of the HPES proton beam via TOF. 
The schematic of LEMS is illustrated in Fig. \ref{fig:LEMS}. 
This setup consists of two LGAD arrays placed directly in the beam path, separated by a flight distance of 40 m.

The size of LGAD is $2\times 2$~cm$^2$ segmented into a $15\times 15$ pixel array, which right match the beam spot size and will maximize the detection efficiency.
The signal from LGAD will be readout by a waveform readout electronics (FROS). 
Since the signal pulse width of LGAD is as short as several nano-seconds, it is required that the sampling rate of the electronics should be as large as 5 Gsps, along with a fitted bandwidth and zero dead time.
The products fulfill this requirement are usually very expensive.
After balance the performance and development cost, the readout of LGAD signals follows the strategy below.
The $15\times 15$ pixel array is bonded in parallel into four readout sectors. 
While this segmentation largely reduced the number of readout channels, it involves a trade-off in timing performance due to increased capacitance \cite{LGAD-4}. 
Laboratory characterization indicates that the time resolution of the bonded LGAD detectors is approximate to 100 ps.

Assuming the two detector arrays possess identical performance, the relative energy resolution of the particle, derived from the TOF method, is governed by the following relation:

\begin{equation}
\frac{\sigma_E}{E} = \gamma(\gamma+1) \sqrt{(\frac{\sigma_T}{T})^2 + (\frac{\sigma_L}{L})^2}
\label{eq:resoE}
\end{equation}

where $\sigma_E/E$ represents the relative energy resolution, $\gamma$ is the relativistic Lorentz factor, and the terms under the square root correspond to the uncertainties contributed by the flight time ($\sigma_T/T$) and the flight path length ($\sigma_L/L$), respectively. 
This equation indicates that the energy resolution is primarily driven by the detector's timing resolution and the geometric precision of the flight distance. 
In the HPES configuration, the uncertainty contribution from the flight path is negligible.
Thus, the energy resolution is dominated by the detector timing resolution. 

\begin{figure}
	\centering
	\includegraphics[width=.9\hsize]{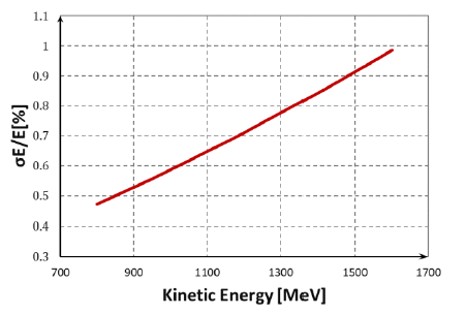}
	\caption{Energy resolution of LEMS, derived with 100 ps timing resolution of LGAD and 40 m distance of flight.}
	\label{fig:LEMS-reso}
\end{figure}

Fig. \ref{fig:LEMS-reso} depicts the calculated energy resolution of the spectrometer as a function of proton energy. 
It is evident that higher proton energies result in degraded resolution due to the reduced sensitivity of velocity to energy in the relativistic regime, thereby placing stricter demands on the timing performance. 
For 1.6 GeV protons, the system is capable of achieving an energy resolution of 1\% considering a timing resolution of 100 ps for each LGAD readout channel.

\subsection{Proton Trigger Device}

\begin{figure}[h]
    \centering
    \subfigure[]{    
		\label{fig:flash_a}     
	    \includegraphics[width=.8\hsize]{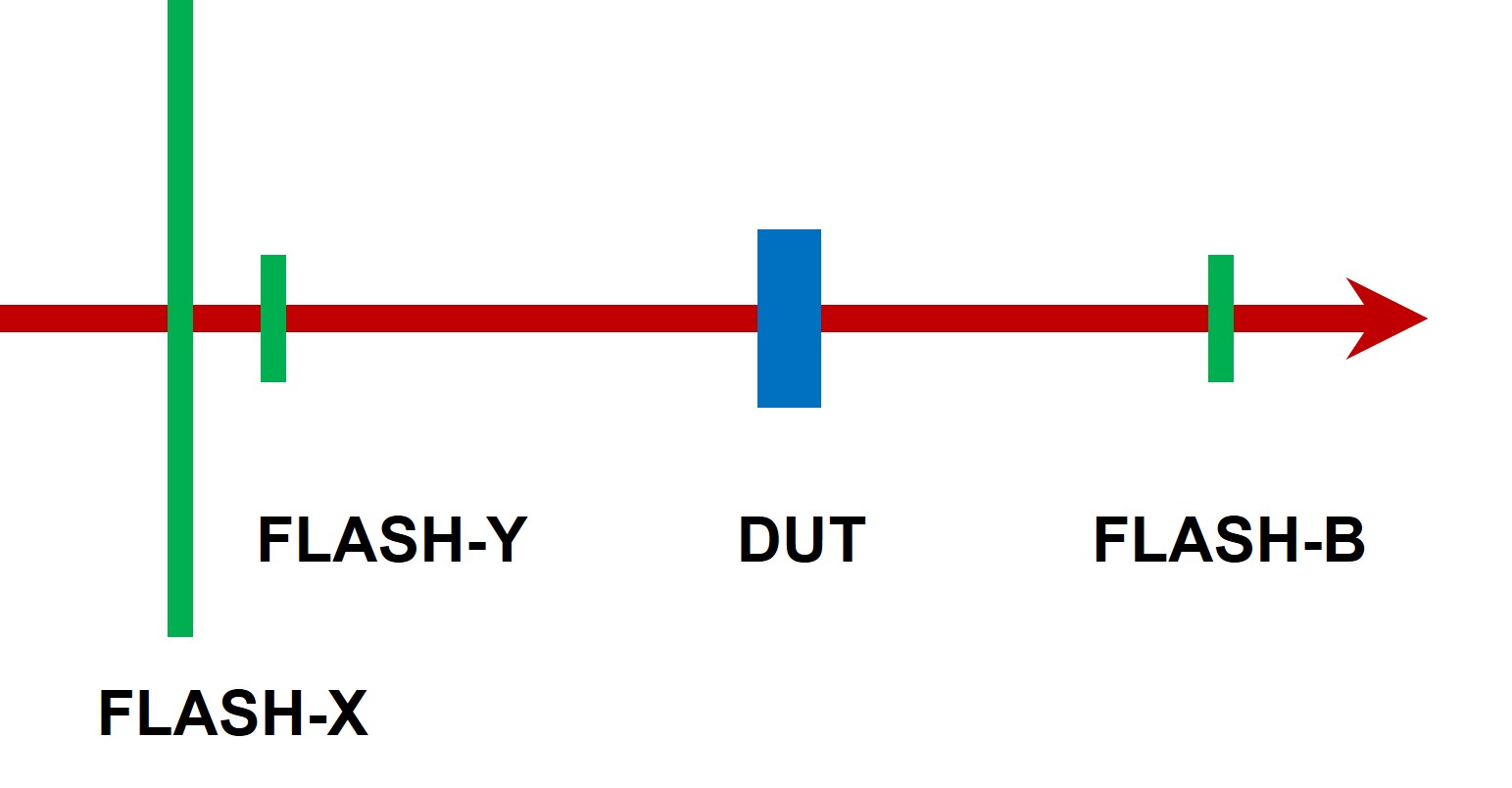}}
    \qquad
    \subfigure[]{    
		\label{fig:flash_b}     
		\includegraphics[width=.6\hsize]{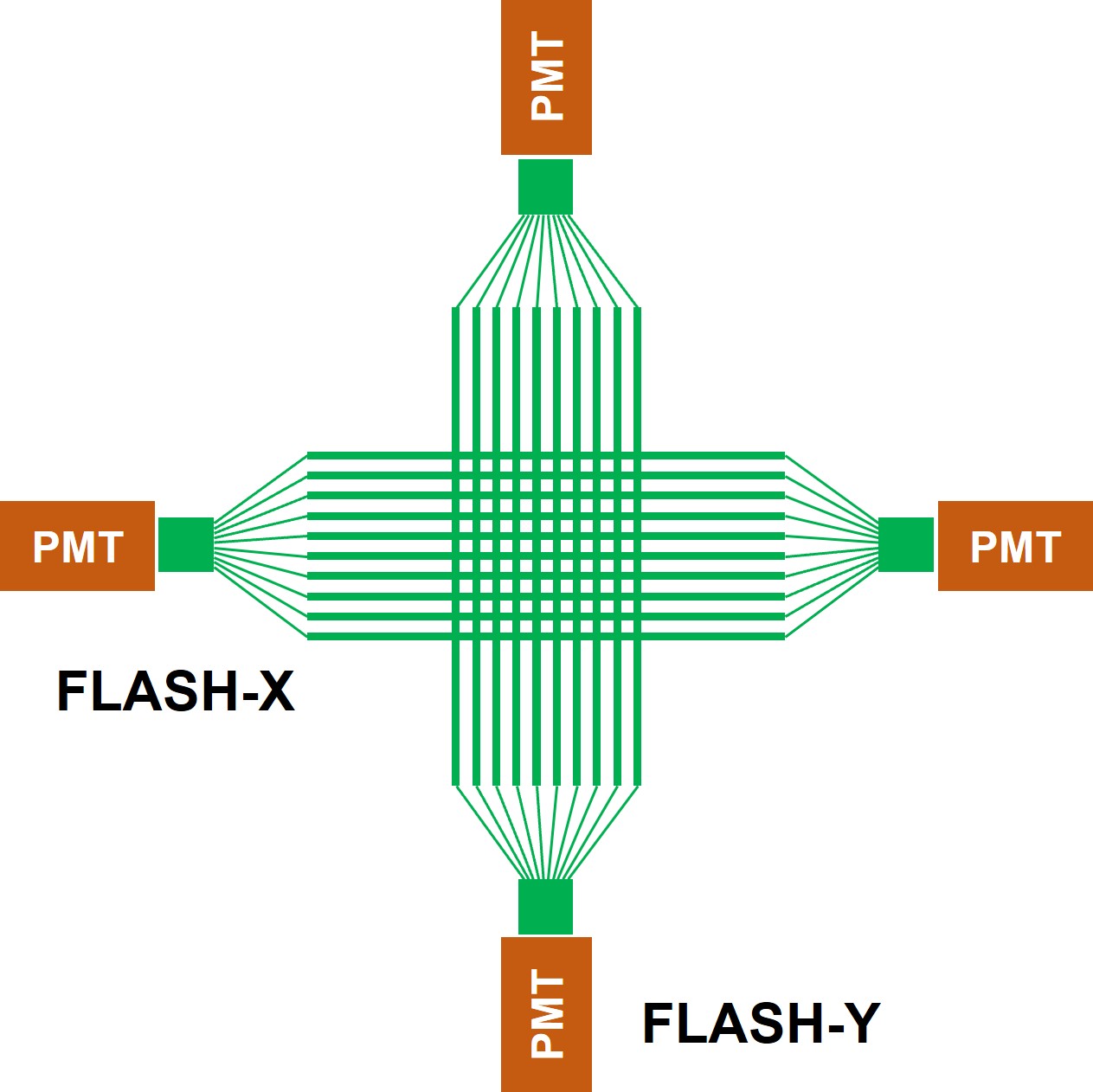}}
    \qquad
    \subfigure[photo of the integrated scintillator fibers]{    
		\label{fig:flash_c}     
		\includegraphics[width=.8\hsize]{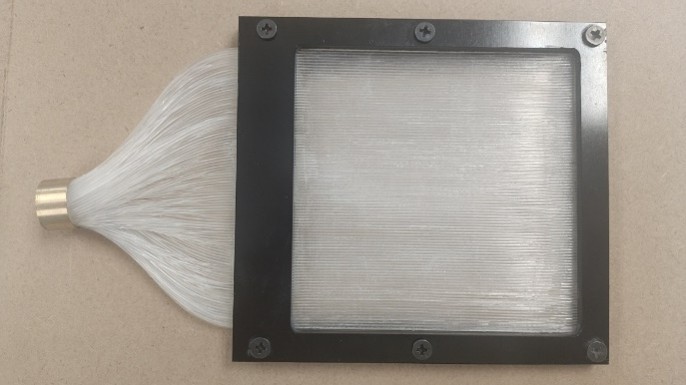}}  
	\caption{Diagrams to the proton trigger device. (a) Deployment of the triple coincidence components along the proton beam. (b) Orthogonal deployment of the upstream components. (c) Photo of the integrated scintillator fibers.}
	\label{fig:flash}
\end{figure}

The HPES incorporates a proton trigger device, named as Fast large Area Starting signal Hub (FLASH), designed to precisely timestamp the proton events and generate the start signal for the data acquisition of detectors. 
By defining a precise temporal gate, this system effectively help to suppress background noise and random coincidences, concerning the 1 ms macro-pulse width and 39 ms macro-pulse blank interval. 
As illustrated in Fig. \ref{fig:flash_a}, the trigger device employs a "2-upstream + 1-downstream" coincidence strategy. 
The upstream component consists of two orthogonal scintillator planes (X and Y directions, refer to Fig.\ref{fig:flash_b}) positioned before the DUT.
This arrangement ensures that triggers are generated exclusively for events originating within the active beam spot, thereby rejecting off-axis background. 
The downstream component involves a third coincidence layer located after the target, which validates that the particle trajectory has successfully traversed the detector stack.

Instead of a scintillator bar, the scintillator fibers are selected as the primary sensing medium (shown in Fig.\ref{fig:flash_c}). 
These fibers are tightly straightened and closely arranged into two layers that are mutually misaligned.
Prompt scintillator photons will be emitted upon energy deposition by charged particles.
The scintillator fibers are converged to a bundle and coupled with PMT.
By adjusting the number of fibers being coupled, the effective area of the trigger detector can be configured as the user's requirement.
In each FLASH component, two PMTs will be coupled with the both side of the fibers and their double-coincidence will suppress the dark-noise.
Preliminary characterization indicates that the system achieves a time resolution of approximately 0.3 ns.

Once the signals generated by the PMT are read by the following front-end electronics, a raw trigger pulse is generated. 
This signal is fast and simple, and will be required by most DUT tests to initiate readout.
More complex tests will necessitate precise data alignment. 
For these cases, a TLU processes the raw trigger pulse to generate a formatted digital trigger signal containing the specific event ID. 
This signal is then fanned out to the respective DAQ systems, ensuring synchronized data collection across the entire experimental setup.
More details about the TLU can be found in chapter \ref{sec:tlu}.

\subsection{Beam Profile Monitor}

\begin{figure}
	\centering
	\includegraphics[width=.9\hsize]{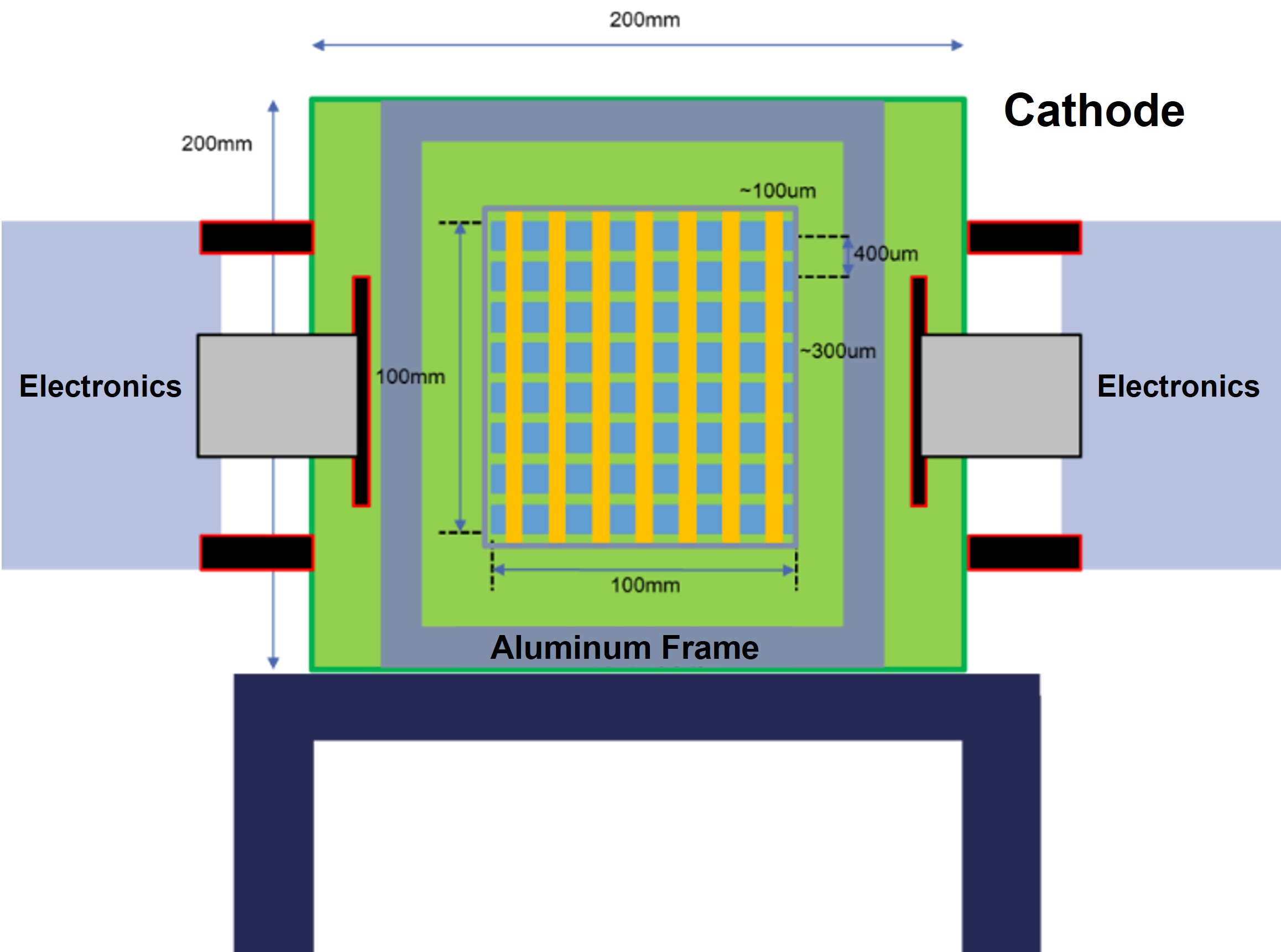}
	\caption{Demonstration to the PALET.}
	\label{fig:pilot}
\end{figure}

A beam profile monitor, named as Proton beAm profiLe dETector (PALET), has been designed for HPES.
The primary function of the PALET is to characterize the transverse spatial distribution of the HPES proton beam. 
The core detector technology applied in the PALET is Micromegas (Micro-Mesh Gaseous Structure) as shown in Fig. \ref{fig:pilot}, which have been widely utilized for large-area particle tracking due to its robust performance and versatility. 
The thermal bonding technology has been implemented in manufacturing the Micromegas for HPES, based on the significant advantages of cost-effectiveness, rapid prototyping, and high production yield. 
Furthermore, Micromegas detectors equipped with resistive anode readout planes exhibit high gas gain, rate capability, and operational stability under high field strengths, making them particularly suitable for beam profiling in high-energy proton environments.


To accommodate varying beam intensities, the system operates in two distinct modes: a low-intensity mode for detector test experiments, covering an active area of $10 \times 10$ to $ 30 \times 30$ mm$^2$, and a high-intensity mode for irradiation experiments, with a scalable area ranging from $10 \times 10$ to $ 100 \times 100$ mm$^2$. 
In both operational conditions, the system is required to maintain a pixel resolution of 300 $\mu$m, which has been determined considering the balance of manufacturing ability and costs.
The resolution has been tested on a 5 GeV electron beam \cite{PALET-USTC}, while the whole system has been well applied on the beam spot profiling on the Back-n white neutron source in CSNS \cite{PALET-CSNS}.

\subsection{Beam Tuning Detector}

The HPES is designed to deliver proton beams across a wide dynamic flux range of $10^3$--$10^8$ protons per second, corresponding to approximately $1$--$10^4$ protons per pulse. 
As previously described, the proton flux is regulated by adjusting the insertion depth of the scattering foil within the main beam of the RCS. 
Consequently, precise characterization of the extracted beam intensity as a function of the scattering foil depth and the collimator aperture is essential. 
Furthermore, in-situ measurement to the extracted flux in terminals is critical to calibrate the instabilities in time structures and flux arising from scattering foil degradation, main beam position drifts, or other systematic variations.

\begin{figure}[h]
    \centering
    \subfigure[Diagram of the PROUD]{    
		\label{fig:proud_a}     
	    \includegraphics[width=.8\hsize]{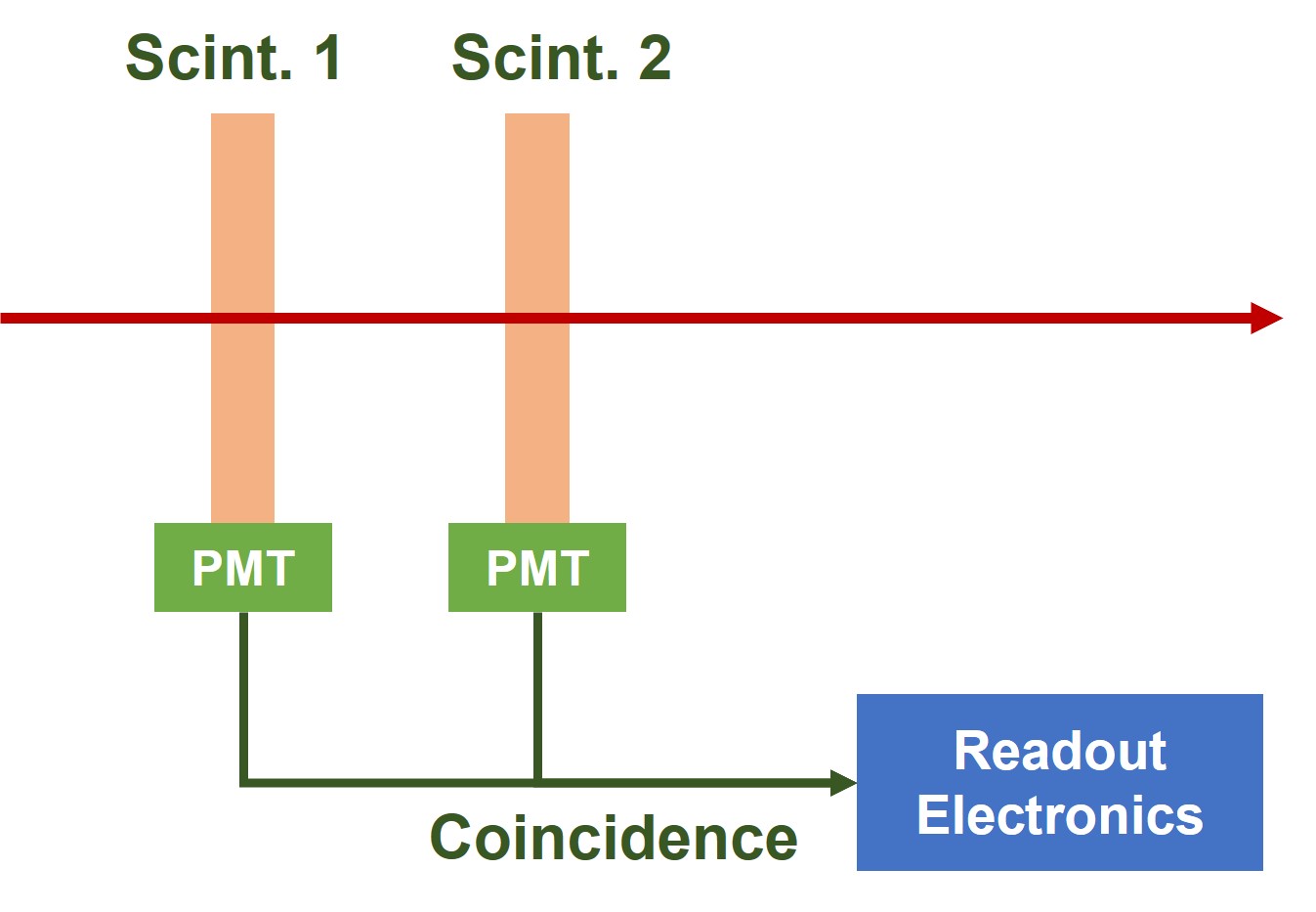}}
    \qquad
    \subfigure[Photo of the scintillator to be used in PROUD.]{    
		\label{fig:proud_b}     
		\includegraphics[width=.8\hsize]{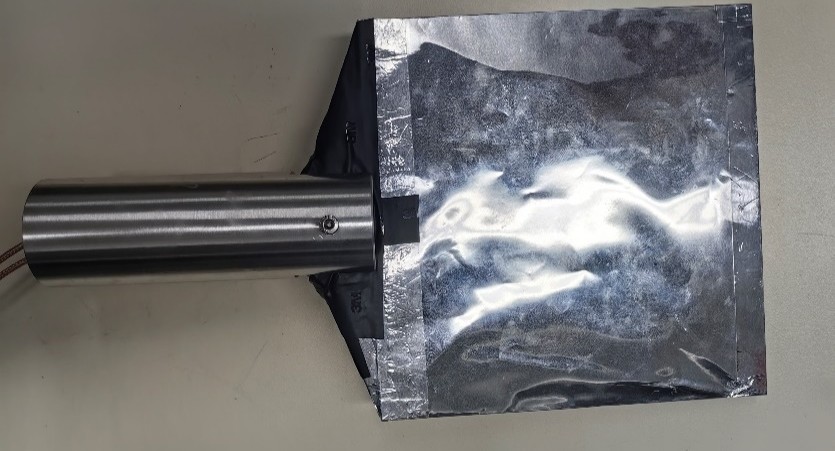}}
	\caption{Demonstration to the PROUD.}
	\label{fig:proud}
\end{figure}

To address these requirements, we have developed the beam tuning detectors named of PROton flUx calibration Detector (PROUD). 
As illustrated in Fig.~\ref{fig:proud_a}, the PROUD system employs two plastic scintillators coupled to photomultiplier tubes (PMTs). 
Signals from both scintillators are processed via a coincidence logic to effectively suppress random background noise. 
Fig.~\ref{fig:proud_b} depicts the mechanical assembly of the encapsulated scintillator and PMT modules. 
The PROUD prototype has undergone performance validation using the muon source at ISIS \cite{ISIS} and cosmic rays in laboratory \cite{PROUD-ISIS, PROUD-CSNS}.
Recent test results indicate that the dynamic range for cosmic ray detection can reach $1$--$8.8 \times 10^3$ protons per pulse \cite{PROUD-CSNS}. 
With further optimization, the system is projected to cover the required dynamic flux range of $10^3$--$10^8$ protons per second.

\subsection{Beam Flux Online Monitor}

\begin{figure}[b]
    \centering
    \subfigure[SEEM]{    
		\label{fig:monitor_a}     
	    \includegraphics[width=.8\hsize]{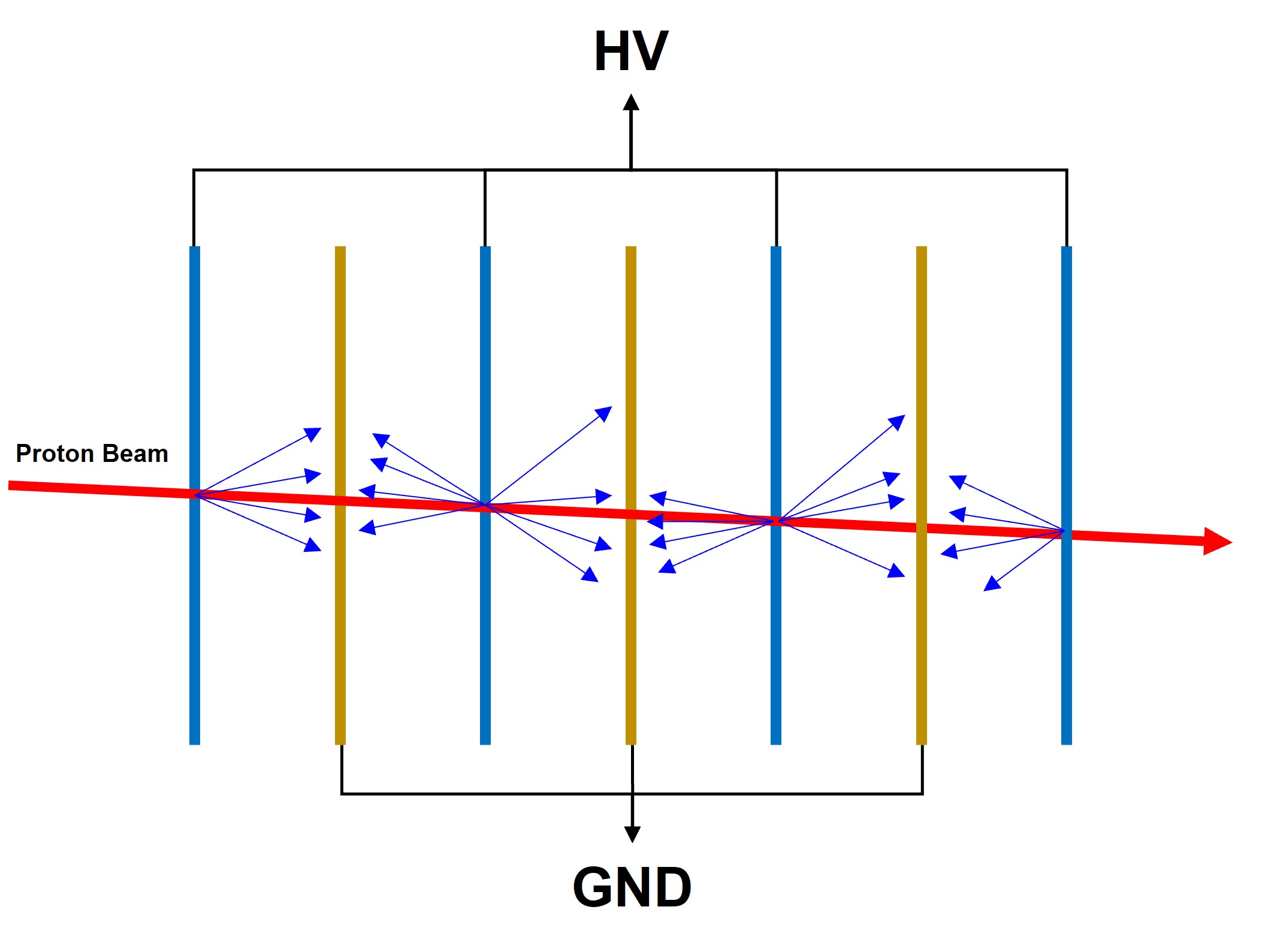}}
    \qquad
    \subfigure[BMOS]{    
		\label{fig:monitor_b}     
		\includegraphics[width=.6\hsize]{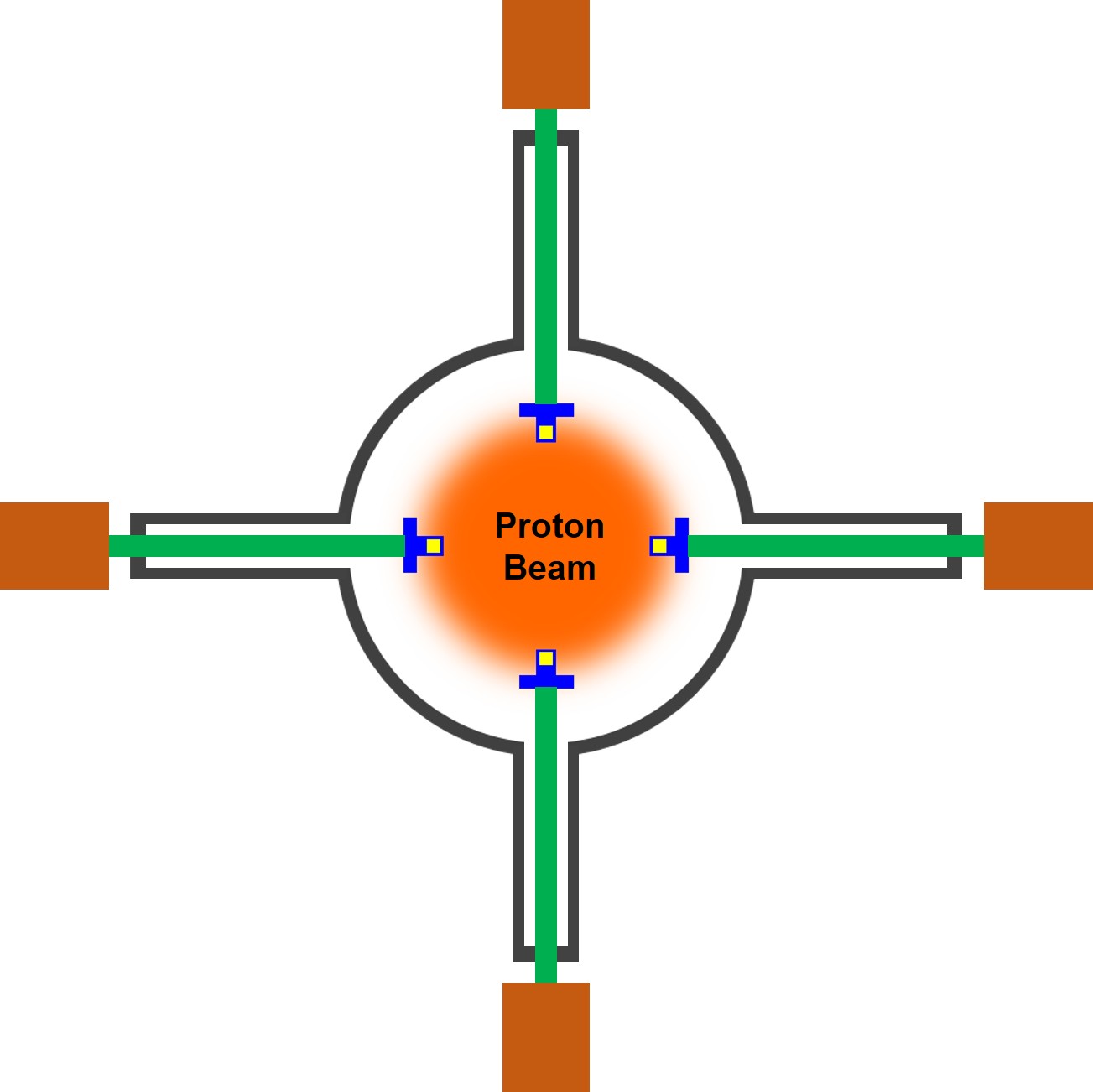}}
	\caption{Diagram of the beam flux online monitor incorporates in HPES.}
	\label{fig:monitor}
\end{figure}

In addition to the requirements for beam tuning and routine calibration, online flux monitoring is essential to provide users with fundamental beam characteristics, along with the beam profile data. 
To address this requirement, two online monitoring devices have been designed for the HPES: the Beam online MOnitor System (BMOS) and the Secondary Electron Emission Monitor (SEEM). 
Unlike the PROUD, both the SEEM and BMOS are deployed upstream of the collimators, thereby providing a direct characterization of the proton flux extracted by the scattering foils.
Within the operation strategy of the HPES, proton flux adjustment is achieved through the coordinated tuning of scattering foils and collimators, with the latter expected to maintain stable performance. 
Consequently, fluctuations in beam flux are primarily attributed to instabilities in the extraction process induced by the scattering foils, such as mechanical malfunctions or rapid beam position drifts in the RCS. 
Therefore, this upstream deployment provides a critical reference for monitoring variations in beam extraction efficiency.

The SEEM, illustrated in Fig.~\ref{fig:monitor_a}, comprises a stack of 10~$\mu$m-thick aluminum foils with an active area of $100 \times 100$~mm$^2$. 
A high voltage bias is applied to the foils in an alternating polarity configuration. 
As protons traverse the foils, secondary electrons are emitted and subsequently collected by the grounded foils. 
The resulting current signals are transmitted to the readout electronics to quantify the proton beam flux.

The BMOS, shown in Fig.~\ref{fig:monitor_b}, employs four Silicon Carbide (SiC) detectors \cite{SiC} that generate discrete signals upon proton hits. 
These detectors are positioned within the beam halo and can be precisely adjusted by servo motors to compensate for slow drifts in the beam center. 
By sampling the particle density in the halo region, the total beam flux can be relatively reconstructed.

\begin{figure*}[]
	\centering
	\includegraphics[width=.8\hsize]{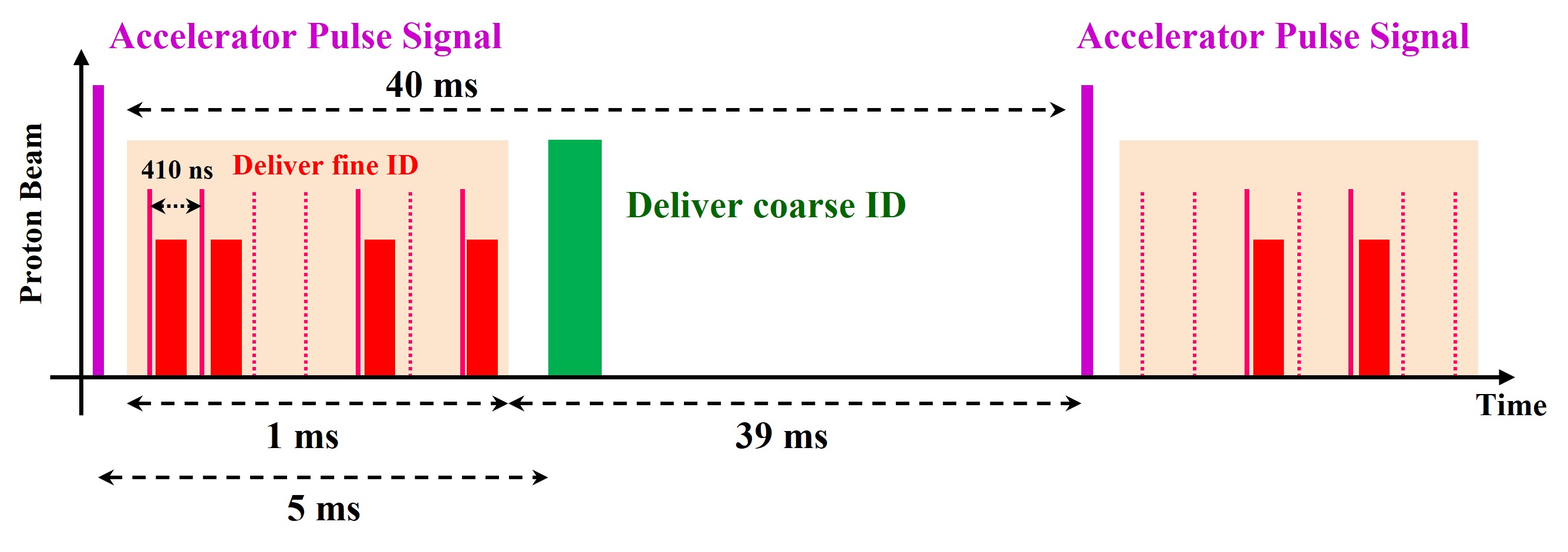}
	\caption{Illustration to the strategy of the dual-section trigger ID, which is composed of fine ID and coarse ID.}
	\label{fig:TLU-dual}
\end{figure*}

\section{Trigger Logic Unit of HPES}\label{sec:tlu}
With the detection devices introduced in Chapter \ref{sec:device}, the basic characteristics and the reference information of protons, including the track and energy, can be provided to the users.
As for the test to the tracking detectors and calorimeters, the alignment of data from DUT to the reference provided by HEPTel and LEMS must be considered to make the data reasonable. 
Thus, the trigger logic unit (TLU) has been developed in HPES to establish the data alignment in level of proton between the DUT and DIT.
This chapter will give an general description to the TLU of HPES. 

\subsection{Motivation to Data Alignment}

Under ideal conditions, data from DUT and DIT, could be aligned in strict chronological order. 
However, in practical beam tests, several factors complicate this alignment. 
First, detectors exhibit varying detection efficiencies.
Discrepancies between the DIT and DUT result in mismatched event sequences in the data storage. 
Second, signal propagation delays from the detectors to the readout electronics introduce timing offsets, which smear the temporal distribution of events within the recorded sequences. 
Third, the presence of "dead time" in certain readout electronics, during which incoming signals are not processed, further increases the difficulty of alignment. 
To address these challenges, a TLU has been developed in HPES.

The TLU is designed to perform the following functions,
It reads output signals from trigger detectors, typically scintillators coupled to PMTs. 
A signal edge crossing a specified threshold is converted into a logic state inside TLU. 
The TLU performs logical operations on these inputs from multiple trigger detectors to generate a unique, ascending Trigger ID, which serves as a label for the proton event. 
This ID is then digitized and distributed in parallel to all devices requiring data alignment. 
These devices subsequently embed the trigger ID into their respective data packages, enabling precise alignment during offline analysis.

Similar systems, such as the AIDA-2020 trigger unit developed in Europe, have been successfully developed. 
By utilizing HDMI interfaces, handshake protocols, and busy-veto mechanisms, the AIDA-2020 delivers digital trigger signals containing a 15-bit Trigger ID, effectively overcoming issues related to efficiency mismatches, timing smearing, and dead time. 
For further details, please refer to \cite{AIDA2020}.

In addition to the TLU solution, data alignment can theoretically be achieved via a synchronous global clock, where a timestamp is broadcast to all devices. 
However, this method is unsuitable for the HPES, as the 410~ns interval between micro-pulses is insufficient for the transmission of high-bit-width timestamps. 
A significant advantage of the TLU architecture is its proven reliability and user-friendliness, demonstrated by its widespread application at major test beam facilities such as DESY and CERN. 
Consequently, the TLU solution was selected for the HPES after a comprehensive evaluation.

\subsection{Design of TLU in HPES}

The TLU design for the HPES is based on the AIDA-2020 architecture. 
Given that the primary objective is to establish data alignment, only the trigger/busy mode (also known as Eudet mode) has been implemented. 
To adapt to the specific beam characteristics of the HPES, several modifications have been made to the original AIDA-2020 implementation.

The first modification concerns the structure of the Trigger ID. 
Under a 40 MHz clock, the transmission of a 15-bit Trigger ID, including essential handshakes, requires approximately 500 ns. 
This duration exceeds the 410 ns interval between micro-pulses. 
Furthermore, a 15-bit ID is insufficient to provide unique labels for all protons within a single experimental run. 
To resolve this, a dual-section trigger ID strategy has been adopted, comprising a fine ID and a coarse ID.

\begin{figure}
	\centering
	\includegraphics[width=.9\hsize]{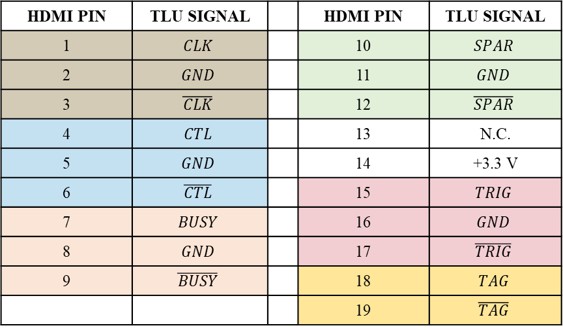}
	\caption{The pin definition concerning the delivery of the dual-section trigger ID.}
	\label{fig:TLU-pin}
\end{figure}

The delivery strategy is illustrated in Fig.~\ref{fig:TLU-dual}. 
The 8-bit fine ID is transmitted during the micro-pulse interval. 
Accounting for handshakes, this transmission requires 325 ns. 
Conversely, the 32-bit coarse ID is transmitted during the macro-pulse interval, which lasts 39 ms and is sufficient for complete transmission. 
In the trigger/busy mode, the final clock pulse is used to initialize the trigger lines.
Therefore, an auxiliary Boolean signal, referred to as "TAG," is provided by the TLU to the DUTs via the remaining pins (as shown in Fig.  \ref{fig:TLU-pin}). 
This differential signal serves as an index: a low level indicates the transmission of the fine ID, while a high level indicates the transmission of the coarse ID.

Given the widespread adoption of the AIDA-2020 architecture at major test beam facilities worldwide, 
the TLU is designed to maintain compatibility with DUTs based on this standard. 
Consequently, DUTs with interfaces developed according to the AIDA-2020 trigger logic communication protocol can operate correctly when receiving trigger signals from the HPES TLU. 
To achieve this interoperability, three key aspects were addressed:

First, regarding the hardware interface, the HPES TLU utilizes the same HDMI connectors and LVDS standards as the AIDA-2020. 
The pin assignments for primary signals remain consistent, with the previously unused pins 18 and 19 enabled to transmit the auxiliary TAG signal. 
For DUTs adhering to the AIDA-2020 architecture, this TAG signal does not interfere with trigger signal transmission in the trigger/busy mode.

Second, concerning digital signal timing, the HPES TLU maintains consistency with the AIDA-2020 in terms of core logic. 
The distinction arises during the Trigger ID readout sequence: a DUT typically sends 16 clock pulses to the TLU. 
The first 15 pulses drive the TLU to transmit a new bit of the trigger ID, while the final pulse initializes the trigger state, with the DUT's BUSY signal asserted throughout this process. 
When interfaced with the HPES TLU, the DUT continues to send 16 clock pulses. 
However, only the first 8 pulses drive the transmission of the Fine ID bits. 
The 9th pulse initializes the trigger state. 
Since the DUT's BUSY signal remains active until the 16th clock pulse is transmitted, the TLU inhibits responses to other proton events during this interval via the BUSY-VETO mechanism.

\begin{figure*}[ht]
	\centering
	\includegraphics[width=.8\hsize]{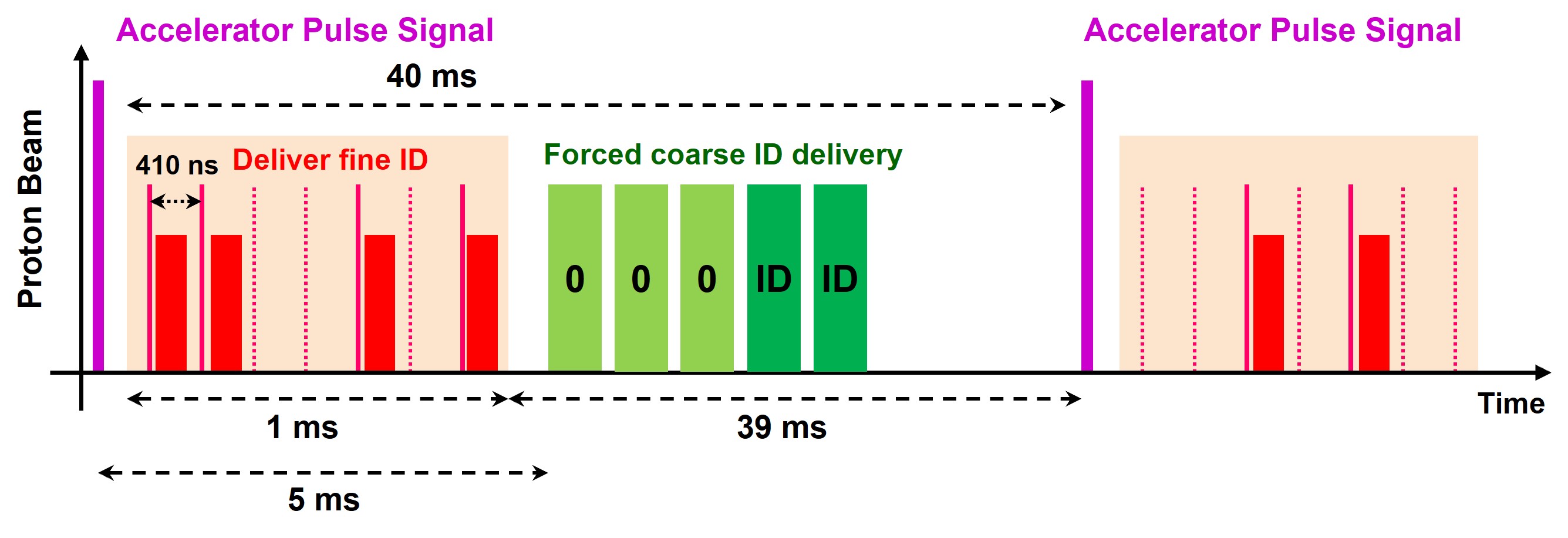}
	\caption{Strategy of delivering the coarse ID to a DUT adhering to the AIDA-2020 architecture }
	\label{fig:TLU-force}
\end{figure*}

Finally, to ensure unique trigger IDs, the HPES TLU supports a forced coarse ID output mode as shown in Fig. \ref{fig:TLU-force}. 
In this configuration, the TLU fans out five complete trigger signals containing trigger IDs during the 39 ms macro-pulse interval. 
The trigger IDs in the first three signals are set to 0, while the concatenation of the trigger IDs in the subsequent two signals constitutes the first 30 bits of the Coarse ID. 
Therefore, users can reconstruct the Coarse ID by applying a logic condition for consecutive trigger ID selection during data analysis.

These three designed features ensure that DUTs adhering to the AIDA-2020 architecture can seamlessly utilize HPES TLU signals without requiring hardware or firmware modifications, necessitating only agreed-upon adjustments to the code of data analysis programs.

\section{Summaries and Prospects}\label{sec:sum}

In this paper, the design of the 1.6 GeV High-energy Proton Experimental Station (HPES) is presented. 
The system extracts a proton beam from the RCS at 1.6 GeV with an energy spread of less than 2.5\%. 
Through the use of a scattering foil, collimator, and degrader, the HPES provides a highly flexible proton beam with an energy range of 0.8–1.6 GeV, a flux spanning $10^3$--$10^{8}$ protons per second, and an adjustable beam spot profile ranging from $10\times10$ to $100\times100$~mm$^2$. 
To facilitate comprehensive user testing, seven devices have been designed for the terminals. 
The trigger device, beam profile monitor, beam tuning detectors and beam flux online monitors are developed to characterize the basic properties of the proton beam. 
Additionally, the proton beam telescope and proton energy spectrometer are designed to provide reference tracks and energy measurements, which are essential for testing particle trackers and calorimeters. 
To address data alignment issues for the DUT, a TLU has been developed, leveraging experience from AIDA-2020 and the specific beam features of HPES.
Compatibility with the AIDA architecture is prioritized to ensure a seamless experience for researchers familiar with the framework.

Beyond detector testing, GeV protons are anticipated to be of significant value in other fields. 
With the rapid development of civil aerospace and artificial intelligence, the exploration of deep space increasingly necessitates on-board supercomputing, intelligent control, and massive data storage. 
Given that GeV protons constitute the dominant component of cosmic rays, validating these technologies requires proton beams in the GeV range. 
Consequently, the HPES is expected to serve as an important platform for investigating irradiation effects on on-board clusters, flight control computers, and data storage devices.
Furthermore, the HPES will contribute to the study of nuclear structure. 
While various nuclear models have been proposed to aid in the design of spallation neutron sources and accelerator-driven sub-critical systems, discrepancies among these models result in significant design uncertainties. 
These uncertainties can be reduced by measuring the multiplicity and double differential cross-sections of spallation products induced by GeV protons. 
Currently, such measurements are limited by a scarcity of data at discrete energy points. 
By providing proton beams with continuous energy, the HPES will enable comprehensive measurements of GeV-proton-induced nuclear reactions, thereby establishing a reliable database for the validation of nuclear structure theories.
The HPES is expected to deliver its first beam by the end of 2029. Upon completion, it will serve as a valuable supplement to global test beam resources and play a pivotal role in the advancement of particle detector technology, aerospace electronics, and nuclear physics.

\end{document}